\def\year{2023}\relax
\documentclass[letterpaper]{article}
\usepackage{aaai22} 
\usepackage{times} 
\usepackage{helvet} 
\usepackage{courier} 
\usepackage[hyphens]{url} 
\usepackage{graphicx} 
\usepackage{graphicx}  
\usepackage{natbib}  
\usepackage{caption}  
\DeclareCaptionStyle{ruled}%
  {labelfont=normalfont,labelsep=colon,strut=off}
\usepackage{algpseudocode}
\usepackage{algorithm}
\usepackage{graphicx}
\usepackage{booktabs}
\usepackage{hyperref}
\usepackage{amsfonts}
\usepackage{setspace}
\usepackage{makecell}
\usepackage [english]{babel}
\usepackage [autostyle, english = american]{csquotes}
\MakeOuterQuote{"}

\usepackage{tablefootnote}

\algnewcommand\algorithmicforeach{\textbf{for each}}
\algdef{S}[FOR]{ForEach}[1]{\algorithmicforeach\ #1\ \algorithmicdo}

\title{Detection and Discovery of Misinformation Sources using Attributed Webgraphs}
\author {
    Peter Carragher, \textsuperscript{\rm 1} 
    Evan M. Williams, \textsuperscript{\rm 1} 
    Kathleen M. Carley \textsuperscript{\rm 1}
}
\affiliations {
    \textsuperscript{\rm 1} Carnegie Mellon University\\
    4665 Forbes Ave, Pittsburgh, PA 15213\\
    \{pcarragh, emwillia, carley\}@andrew.cmu.edu
}

\begin{document}
\maketitle
\begin{abstract}

Website reliability labels underpin almost all research in misinformation detection. However, misinformation sources often exhibit transient behavior, which makes many such labeled lists obsolete over time. We demonstrate that Search Engine Optimization (SEO) attributes provide strong signals for predicting news site reliability. We introduce a novel attributed webgraph dataset with labeled news domains and their connections to outlinking and backlinking domains. We demonstrate the success of graph neural networks in detecting news site reliability using these attributed webgraphs, and show that our baseline news site reliability classifier outperforms current SoTA methods on the PoliticalNews dataset, achieving an F1 score of 0.96. Finally, we introduce and evaluate a novel graph-based algorithm for discovering previously unknown misinformation news sources.

\end{abstract}

\section{Introduction}


In 2014, Sinclair Treadway and Sean Adl-Tabatabai founded YourNewsWire, which over time was identified by many fact-checkers as a purveyor of misinformation. In 2019, the co-founders, blaming loss of revenue on Facebook's fact-checking system, rebranded their website by simply migrating the domain name from "yournewswire.com” to “newspunch.com" \citep{mash2019fake}.

The detection of misinformation sources among news domains relies heavily on preexisting domain reliability labels. However, these curated domain lists are subject to two core limitations. Firstly, as illustrated above, building automated systems to detect websites that consistently spread misinformation is made challenging by the ease of the evasion and exit tactics available to bad actors. As researchers and fact-checking organizations identify and release misinformation sources, the owners can shut down the site, migrate to a new domain, or simply start over with a new website. Consequently, we observe that over 50\% of domains on unmaintained blocklists published between 2017 and 2019 are dead. This presents a major obstacle for misinformation source detection and discovery tasks.

Secondly, the vast majority of misinformation source research relies on signals mined from public social media data. As misinformation becomes increasingly prevalent in developing countries where social media usage remains uncommon or where English is not the primary language, detection and discovery methodologies that are not dependent on website content or social media are needed \citep{okereke2021covid}.

We attempt to address these obstacles by demonstrating the predictive power of webgraph and Search Engine Optimization (SEO) data in detecting news site reliability and bias. This paper does not rely on social media data to identify unreliable domains, but rather relies on pointers between websites. We provide three primary contributions in this work. Firstly, we introduce a novel attributed webgraph dataset with labeled news domains and their connections to top outlinking and backlinking domains, i.e. domains to which a target domain links and domains that link to the target domain. Secondly, with extensive experiments, we demonstrate the predictive power present in the data by using graph neural network (GNN) models to predict the reliability and bias of reliable and unreliable news domains. We show that these models outperform existing work on the PoliticalNews dataset \citep{page_detection_topic_agnostic}. We make all data \footnote{https://doi.org/10.1184/R1/25174193.v1} and code public \footnote{https://github.com/CASOS-IDeaS-CMU/Detection-and-Discovery-of-Misinformation-Sources}.

Finally, we present a novel webgraph-based discovery approach that is not dependent on language, content, or social media data, and we demonstrate the success of the approach in identifying previously-unknown unreliable news domains. The content-agnostic and social media-agnostic nature of this work differentiates our approach from all previous work on this topic and provides several key advantages. Firstly, this allows our approach to be implemented in non-English settings. Secondly, our approach is not reliant on API access to social media data. Finally, our content agnostic system will not be impacted by increasing use of generative AI in SEO settings, which has been identified as an emerging threat \citep{page_detection_time}.


\section{Related Work}

\subsection{Misinformation Detection}


Classifying individual texts or articles is a core problem addressed by research in misinformation detection. Castelo et al. address misinformation page detection using article content \citep{page_detection_topic_agnostic} which Chen and Freire extend to unreliable domain discovery using social media context \citep{domain_discovery_social_media}. Building on that work, \citet{domain_discovery_sm_page_detection_sm_content} combine website content and social context resulting in a misinformation page detection method that leverages multi-modal data. 

We depart from the article and content-centered approaches to misinformation detection in that our detection systems are at the domain-level. While a single article may be considered misinformation, the collection of articles published by a single news domain has an associated probability distribution that captures how likely an article is to contain misinformation, given the source. In content-centered approaches, reliability scores are assigned to domains based on this distribution, labeling domains that frequently publish articles containing misinformation as "unreliable". As we are concerned with misinformation \textit{sources}, we discuss unreliable domain detection rather than misinformation \textit{page} detection, using `misinformation source' and `unreliable domain' interchangeably.

Motivating our domain-level and content-agnostic approach, \citet{page_detection_time} employed change-point detection to reveal that the linguistic features outlined in \citet{page_detection_topic_agnostic} evolve over time. They find significant variance in the psychological features that distinguish reliable articles from those peddling misinformation, making unreliable page detection difficult over long periods of time.  Our webgraph methods present a novel approach to unreliable domain detection and discovery which avoids common pitfalls of the content-based approach.

\subsection{Webgraphs}

Several studies have explored the predictive power of webgraphs on tasks related to unreliable domain detection. Using community detection methods, \citet{www_bias_detection} scrape links from US and UK news sites and find clusters based on political bias using labels from Media Bias Fact Check (MBFC) \citep{mbfc}. Similarly, \citet{hrckova2021quantitative} found that partisan news domains in central Europe form clusters in webgraph data. \citet{link_scheme_misinfo} present a case study on hyperlink and social media usage for peddling misinformation. They explore coordinated use of hyperlinks and social media on a dataset of 1.4k URLs with reliability labels drawn from various sources, including MBFC. Lee et al. used the linkage patterns of labeled news domains shared by Twitter users to visualize connections of domains to unreliable domains on Twitter \cite{lee2022explaining}. These analyses provide theoretical justifications for using webgraph data to detect website reliability and bias. 

\subsection{Search Engine Optimization}
SEO is an 80 billion dollar industry built to increase traffic to websites \citep{seo80bil}. High-ranking pages are far more likely to be seen. In a 2013 analysis of 300 million search engine clicks: 92\% were on first page of search results with the majority of clicks going to just the top two \citep{insights2013value}. Furthermore, users were found to be 140\% more likely to click the last result on the first page than the first result on the second page \citep{insights2013value}. Consequently, SEO has become ubiquitous. On a study of over 250k search results, \citet{seo_prevalence} found that only 12\% were likely not to have been optimized, and 58\% were certainly optimized. \citet{social_media_vs_link_building} contrast the effectiveness of SEO and social media-sharing for media promotion. They find that while social media-sharing results in more immediate boosts to traffic in the short term, link-building is superior in the long term. In this work we elect to focus on link building, leaving the role that social-sharing strategies plays in boosting unreliable new sources to future work.


\paragraph{Black-hat SEO} While many SEO practices are endorsed by search engines, those aimed at gaming recommendation algorithms are forbidden by search engine webmaster guidelines\footnote{\href{https://developers.google.com/search/docs/essentials/spam-policies}{See Google's spam policies for web search.}}---these tactics are typically referred to as "black-hat" SEO. For a detailed overview of white-hat and black-hat SEO methods, we refer the reader to the work done by \citet{black_hat_seo_review}. Common black-hat SEO techniques of relevance here are: (1) attracting search engine indexers through networks of blogs (`Blog Ping'), (2) automated user engagement in forums or comment sections (`User Generated Content'), and (3) link building by spamming (`Link Scheming'). More recently, \citet{black_hat_seo} analysed the sentiment of SEO service customers and found that 32\% of respondents believed that the widespread use of black-hat SEO methods by SEO service providers was damaging their ranking. \citet{link_scheme_detection} suggest that clustering domains based on high level SEO attributes such as Pagerank score and Domain Authority can identify black-hat link scheme sites. This motivates our approach to identifying link scheme sites, which is the first step in our network-based unreliable domain discovery process (Algorithm \ref{alg:link_scheme_identification}).

\paragraph{SEO and Misinformation} Search engine rankings have broad impacts even outside of e-commerce; it has been shown that Search engine rankings can have a substantial impact on the political beliefs and voting patterns of users \citep{SEO_elections}. Three laboratory experiments with double-blind control group design found that relatively minor changes in search engine rankings could influence decisions of undecided voters \footnote{The magnitude of the effects in this work are disputed by \href{https://algorithmwatch.org/en/watching-the-watchers-epstein-and-robertsons-search-engine-manipulation-effect/}{Katharina Zweig, who finds that the cumulative effect is likely on the order of 2-4\% rather than the 20\% reported in the study.}}. Search engine audit studies---studies where researchers query terms on multiple search engines and compare the rankings---have revealed a link between SEO and misinformation. For example, \citet{urman2022earth} executed conspiratorial searches on Google, Bing, DuckDuckGo, and Yandex and found that Yandex and DuckDuckGo consistently return misinformation from unreliable sources in their top search results. \citet{bradshaw2019disinformation} analyzed how a set of 29 junk news sites optimize keywords to increase web-traffic and spread disinformation over a three year period. More recently, \citet{williams2023search} demonstrated the usefulness of webgraph data in identifying large-scale link scheme activities around Kremlin-aligned propaganda domains. We build on these efforts in an attempt to link specific black-hat SEO tactics to unreliable news sources.

\subsection{Domain Monitoring}

\citet{censor_lists} find that older website blocklists have a high rate of parked domains---domains that are up for sale and not currently in use. Domain survival rates are estimated using a classifier that is trained to flag parked domains \citep{parking_classifier, regexp_parked}. 
We note that this limitation of blocklists has \textit{not} been recognized by previous work on unreliable domain lists.

\section{Tasks}


We evaluate our approach using three classification tasks. 1) \textit{Reliability Prediction}: a binary classification task determining domain reliability. 2) \textit{Relative Political Bias}: binary "left"-"right" partisan lean classification. 3) \textit{Absolute Political Bias}: a binary "center"-"extreme" task determining whether the site publishes extremist or centrist content. 


Previous research and government reporting has demonstrated how IRA trolls amplify both extreme left and extreme right sources to sow discord and increase polarization \citep{bradshaw2022playing}. Training separate classifiers for left-right and extreme-center dimensions of political bias enables a more fine-grained discovery approach. 

Given the transient nature of many misinformation domains, classification of existing labeled domains may be of limited value. We therefore introduce a final task, 4) \textit{Misinformation Domain Discovery}, to allow us to identify new misinformation domains. We split this process into two conceptual steps: 1) Develop a set of criteria for identifying suspicious backlinking patterns and 2) Extract unreliable websites to which suspicious websites link.

\section{Data}

\subsection{News Site Identification}
To identify a starting set of news sources for our tasks, we leverage information from two different sources. The bulk of our domain list is made up of domain labels scraped from MBFC\footnote{\url{https://mediabiasfactcheck.com/methodology/}} \citep{mbfc}. MBFC has six grades of reliability, ranging from very low to very high. MBFC also provides political bias labels ranging from extreme left (-2) to extreme right (+2), the distribution of which is given in figure \ref{fig:bias_by_reliability}.

Domains collected by \citet{blacklist} were part of an investigation into the role of social media in spreading misinformation during the 2016 US Presidential Elections. As part of their analysis, they publish lists of unreliable sites from various investigative sources; Buzzfeed, Politico, and Snopes.com, as well as domains identified in their own analysis. They also draw on a blocklist built by \citet{original_list}, which published a dataset of 125 URLs extracted from articles scraped from Politico, Snopes.com, and Facebook. URLs are extracted based on keywords pertinent to the 2016 election. We combine the MBFC data with these previous lists \citep{blacklist} for a total of 4,206 domains. A recent study on the correlation of news reliability ratings constructs a similar list \citep{domain_rating_correlation}. Table \ref{tab:survival_rates} gives a count of domains per source.



\subsection{Domain List Analysis}

We investigate the survival rates of domains from previously-published unreliable domain blocklists, and filter out dead domains. For each domain we determine whether the domain responds to GET requests and whether the domains are parked, i.e., are not being held for sale by a parked domain registrar. If a GET request returns a 404, the site is dead. If the domain is parked or dead, then it should no longer be included in the dataset.

To determine whether domains were parked, we build a boosted decision tree classifier following the methodology proposed in \citet{parking_classifier}. For more information on the Parked Domain classifier methodology, see the appendix. When applied to our list of domains, the model identified 248 parked domains with 96\% precision (based on a manual evaluation on a 10\% sample). 



In Table \ref{tab:survival_rates}, we show that the older labeled sources from \citet{blacklist} have low survival rates. MBFC is continuously-maintained, so it has high survival rates. With the exception of the Snopes.com list (which is drawn from the dedicated fact-checking site), the non-maintained lists have survival rates of less than 50\%. The number of links that point to a site is also an indication of the status of a site  \citep{ahrefs}. We observe that simply dropping news websites with less than 10k backlinks excludes the majority of the dead domains: it filters out 63\% of domains returning 404s and 87\% of parked domains. We postulate that this makes filtering by backlinks a reasonable heuristic for the parked domain classifier and the 404 classifier in unreliable domain discovery. This further simplifies our method such that parked domain and 404 classifiers are not needed.

In summary, we filter out any sites that have less than 10k backlinks as those are unlikely to be influential news sites. Our final domain reliability dataset contains 3,211 labeled domains which we will refer to as \textbf{MBFC*} (\autoref{tab:survival_rates}). 

\begin{table}[!h]
  \begin{tabular}{cccccccl}
    \toprule
    Source&URLs& !404 & !Parked & Both &$>$10k&  \\
    \midrule
    MBFC	    & 3227         & 97\%	     & 99\% & 96\% & 87\%	 \\
    MBFC-Q	    & 426           & 90\%	     & 94\% & 84\%& 67\%  \\
    Snopes	    & 55	         & 80\% 	     & 84\% & 67.5\%& 69\% \\
    Blocklist	& 359   	       & 74\%	     & 59\% & 43.5\%& 11\% \\
    Politico	& 83             & 65\%	     & 68\% & 44\%& 18\% \\
    Buzzfeed	& 55	         & 70\%	     & 50\% & 35\%& 11\% \\
    \midrule
    Total       & 4205 & 3910 & 3952 & 3658 & \textbf{3211} \\
  \bottomrule
\end{tabular}
\caption{Unreliable Source Survival Rates}
\label{tab:survival_rates}
\end{table}

\subsection{MBFC* Site Labels}
To unify labels from each source, we needed to binarize labels. There are some sites that MBFC classifies as `questionable', but it does not assign these sites a reliability score\footnote{\url{https://mediabiasfactcheck.com/fake-news/}}. After inspecting a sample of these domains, we chose to assign these questionable websites an unreliable label. We handle overlap between lists by taking the most recent label available. This results in current MBFC labels having precedence over the labels assigned by earlier works. 

The blocklist domains label unreliable websites as black (very low reliability), red (low), and orange/yellow (mixed), which we map to the corresponding labels used in MBFC. Remaining consistent with \citet{domain_discovery_social_media}, any domain with a reliability rating of mixed or lower is assigned an unreliable label, the rest are assigned reliable labels.

For the bias tasks we use only the labels from MBFC as the blocklist sites do not have bias labels \citep{blacklist}). Using the $>$10k backlink criteria for filtering out dead domains, we are left with 2,629 bias labels. For the relative political bias task, we label the domain left if the MBFC bias score is less than 0, right if the bias score greater than 0. Labels of value zero, i.e., centrist labels, are dropped for this task. For the absolute bias task, bias scores of -1, 0, and 1 are assigned the centrist label; otherwise we label it extreme. 

\begin{figure}[!h]
    \centering
    \includegraphics[height=6cm]{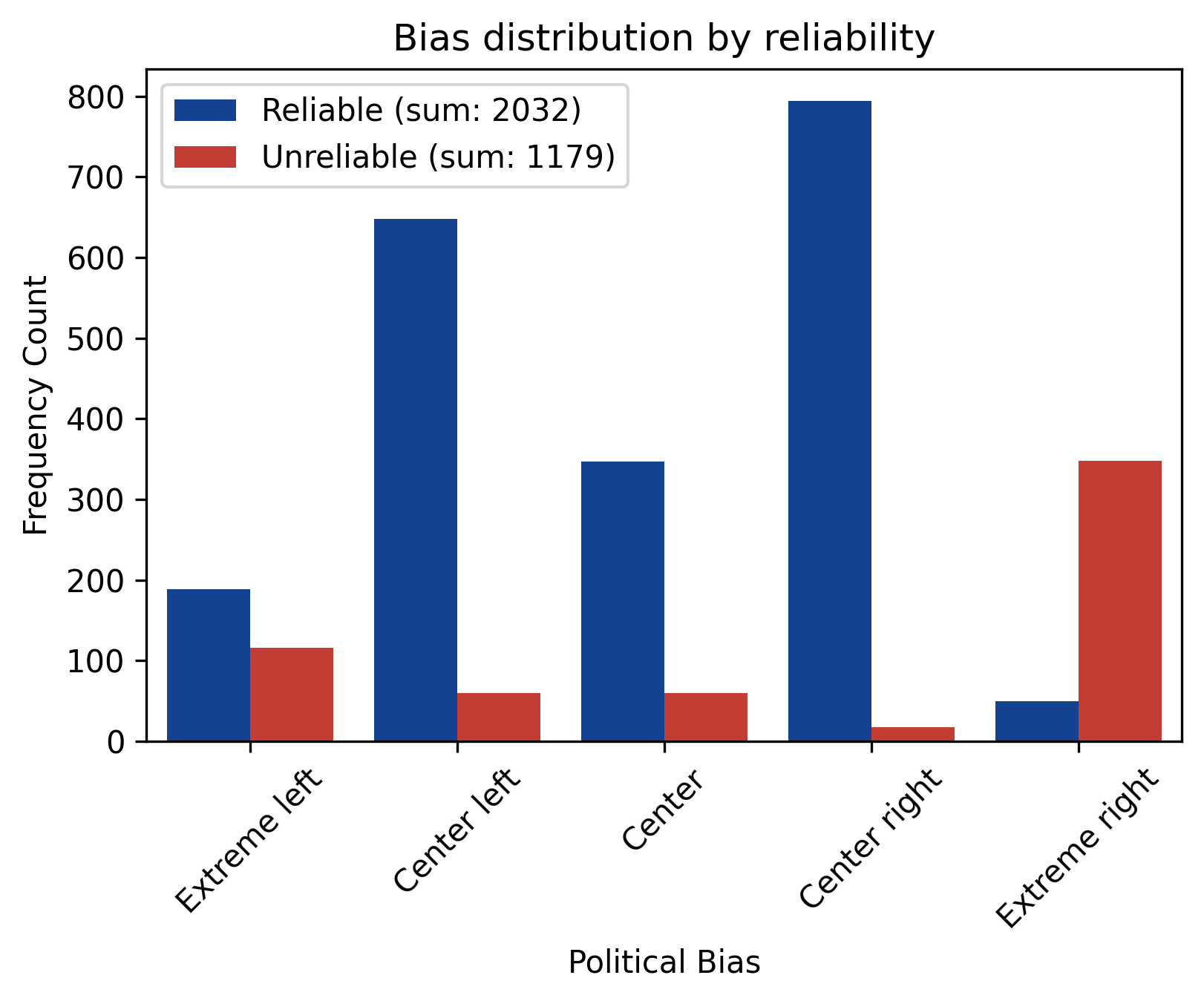}
    \caption{Bias label counts grouped by reliability.}
    \label{fig:bias_by_reliability}
\end{figure}

Our two final datasets contain 3,211 reliability-labeled domains and 2,629 bias-labeled domains. In Figure \ref{fig:bias_by_reliability} we visualize the distributions of political bias labels for each news site reliability label. Aside from a 2:1 ratio of reliable to unreliable domains, we note reliable domains tend to to skew centrist, whereas unreliable domains tend to skew towards political extremes. Exploratory data analysis shows that SEO attributes align with our intuition with respect to website reliability (see \autoref{fig:edu} in the appendix). For example, we find a positive correlation between the binary reliability labels and the proportion of backlinks that come from .edu domains ($r = 0.09$), as well as from .gov domains ($r = 0.08$). 

\subsection{PoliticalNews} 
The PoliticalNews dataset is a popular evaluation dataset in fake news detection studies. It was proposed by \citet{page_detection_topic_agnostic} alongside the Topic-AGnostic (TAG) classifier for use in misinformation page detection. It has since been adopted by \citet{domain_discovery_social_media} to evaluate unreliable domain detection methods. It consists of 137 domains, including 58 news domains and 79 unreliable domains collected between 2013 and 2018. As other detection methods have published results on this evaluation set \citep{page_detection_topic_agnostic,domain_discovery_social_media}, it serves as a useful benchmark for our SEO-based detection method.

\subsection{Site Attributes}
We pull SEO attributes for both MBFC* and PoliticalNews using Ahrefs---a proprietary search engine optimization toolkit that purports to possess a 13 trillion link database and the second most active commercial webcrawler after Google\footnote{\url{https://ahrefs.com/robot}}. We pull 24 different attributes for each domain in our dataset, including the number of .edu domains that link to the website, total number of backlinks, and whether or not the domain has links from user-generated content (\autoref{fig:feature_importances}).
 
\subsection{Constructing a Webgraph}

\begin{figure}[!h]
    \centering
    \includegraphics[scale=0.18]{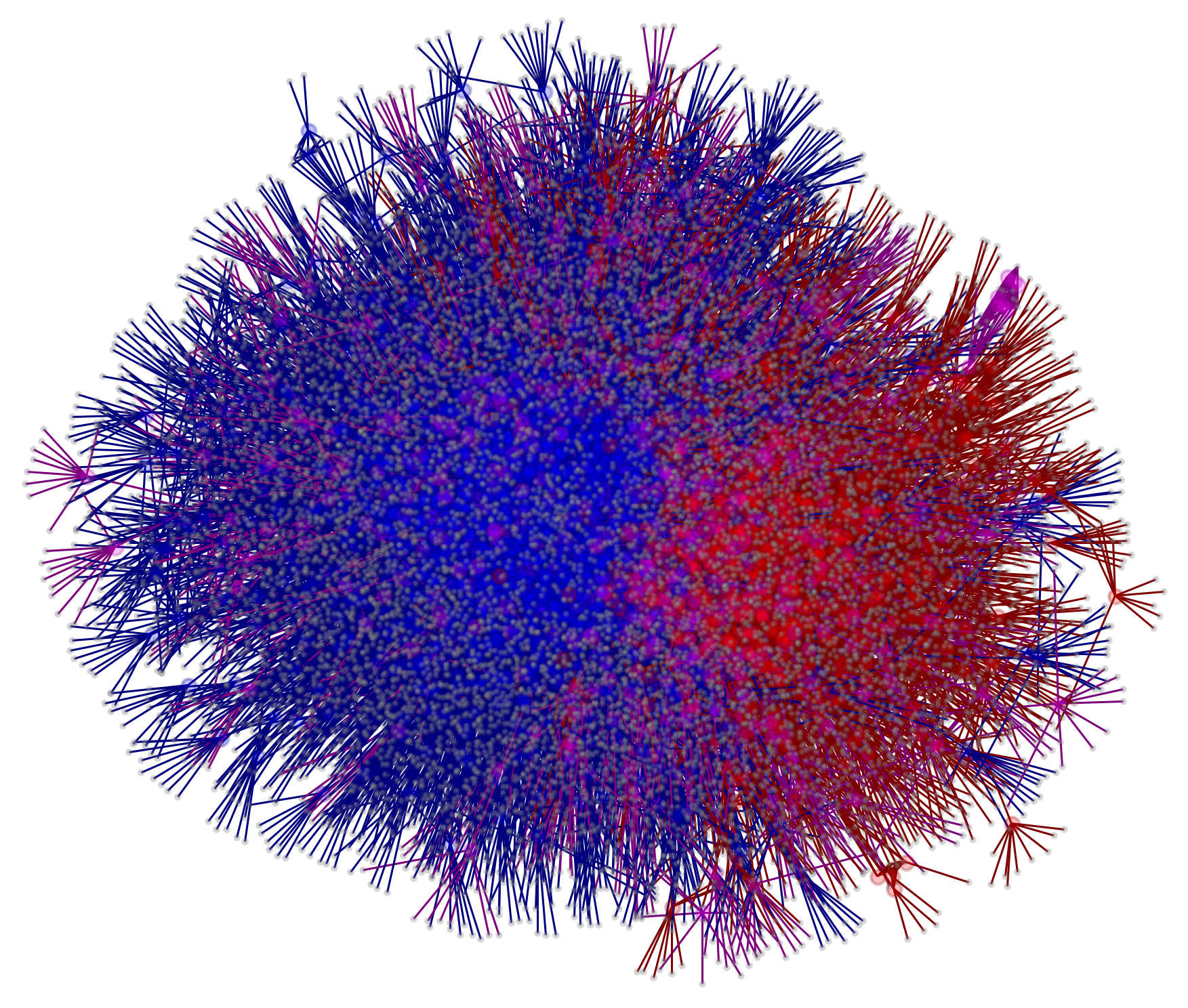}
    \caption{Backlink network where node colors show reliability labels; red are low reliability, blue are high reliability, pink are mixed reliability, grey are backlinking domains.}
    \label{fig:backlink_network}
\end{figure}

We construct webgraphs for the MBFC* dataset using top backlink and outlink domain data from Ahrefs. As a result of API and monetary constraints, we limit the webgraph to a single hop. We extract three networks for each labeled domain: a backlink network (top 10 backlinks per news site), an outlink network (top 10 outlinks per news site), and a combined network (the union of the top 10 backlink and outlink networks). Once data were pulled, we constructed weighted and attributed Graphs (webgraphs), $\mathcal{G} = {V, E}$ where $V$ are websites and $E$ are represent either backlinks or outlinks of domains. We visualize the backlink network colored by news-site reliability in Figure \ref{fig:backlink_network}. Visually, unreliable and reliable domains seem to largely cluster together. This visual separability suggests that we should be able to use network signal to classify reliability with reasonable accuracy. For summary statistics on the three networks, see Table \ref{tab:summary_stats} in the appendix. 

\section{Misinformation Source Detection}


\subsubsection{Flat Baselines}
For our flat baseline experiments we use only the SEO attributes for each URL, ignoring network structure as well as the top backlinking and outlinking domains. We then run standard flat ML baselines: gradient boosted decision tree (GBDT), random forest, decision tree, MLP, and SVM. For parameters we use 50 estimators for the random forest classifier, 2 hidden layers of dimension 200 for the MLP and the linear kernel for the SVM. Scikit-learn defaults are used for the remaining parameters. We repeat these baselines for the reliability, absolute political bias, and relative political bias tasks on the labelled dataset. Five-fold cross validation is used with an 80:20 training-test set split.

\subsubsection{GNNs}
For our GNN experiments, our goal is to leverage local homophily present in our partially-labeled SEO network in order to better classify unreliable and biased domains. Formally, for a given task, for website $u$, our goal is to create node embeddings $z_u \in \mathbb{R}^d$ that map $u$ to its corresponding one-hot-encoded label $y_u \in \mathbb{Z}^c$. The layerwise propagation of a Graph Convolutional Neural Network (GCN) \cite{kipf2016semi} can be written as:
\begin{equation}
H^{(l+1)} = \sigma(\tilde{D}^{-1/2}\tilde{A}\tilde{D}^{-1/2}H^{(l)}W^{(l)})
\end{equation}
where $\sigma$ is an activation function, $\tilde{A}$ is the sum of the Adjacency matrix $A$ and its identity matrix $I$, $\tilde{D}$ is the diagonal degree matrix of $\tilde{A}$, $H^{(l)}$ is the $l^th$ layer activation matrix, and $W^{(l)}$ is a trainable weight matrix.

For each task, we implement a homogeneous two layer GCN with 64 output channels that feed into a final linear layer with relu activation and a dropout of 0.5. We found that using GCN layers outperformed GraphSage \cite{hamilton2017inductive} and GAT \cite{velivckovic2017graph} layers on our baseline architecture, so we chose to use GCN layers for all subsequent experiments. We use crossentropy loss function and and an Adam optimizer. A log softmax activation is applied to the final layer.

The graph is partially labeled as many backlinks and outlinks do not have reliability or bias labeled. Consequently, unlabeled nodes are masked, but their features are still available to labeled neighbors during the propagation step. To train, we use a transductive random 80:10:10 split and 0.05 for the learning rate. We select models using early stopping with a patience of 30 and a minimum delta of 1e-4.

\paragraph{Link Normalization} We investigate several link normalization schemes, as GNN-based models have been shown to work better when edge weights are normalized. The link data can be represented as an Adjacency Matrix $A$ where $A_{i,j}$ represents the number of links from domain $i$ to $j$, and the (diagonal) \textit{backlink} degree matrix $D_b$, and \textit{outlink} degree matrix $D_o$, where $D_{i,i}$ represents the backlink and outlink total for domain i respectively. As we only observe links from the 10 highest-volume backlinking and outlinking domains, we also have the sampled degree matrices $\hat{D_b}$ and $\hat{D_o}$, with diagonal elements representing the sum of all incoming and outgoing links observed \textit{in the graph}. We explore 7 different edge-weighting approaches:

\begin{itemize}
    \item \textit{links}: $A_{i,j}$
    \item \textit{log links}: log($A_{i,j}$) iff $A_{i,j} > 0$
    \item \textit{backlink}: $D_b^{-1}A$ --- i.e. the \% of j's backlinks from i 
    \item \textit{outlink}: $AD_o^{-1}$ --- i.e. the \% of i's outlinks that go to j
    \item \textit{graph-backlink}: $\hat{D_b}^{-1}A$
    \item \textit{graph-outlink}: $A\hat{D_o}^{-1}$
    \item \textit{page}: \% of j's reference pages from i's domain
\end{itemize}

Backlink and outlink edge weightings of the form $AD^{-1}$ and $D^{-1}A$ follow the random-walk normalization method. We also include \textit{page}, an edge weight calculated on reference page count---the number of unique pages on a domain that link to a target domain. Intuitively, when \textit{page}$_{i,j}$ is near 0, links from domain $i$ to $j$ come from a narrow subset of domain $i$ subpages relative to the total number of webpages that reference domain $j$.

\subsection{Detection Results}
Given the label imbalance on the MBFC* unreliable classification task, we evaluate models using Binary F1 where "unreliable" is the positive label (\autoref{tab:baseline_experiments}). 

On the PoliticalNews domain list, we replicate the TAG classifier using data released by the authors \citep{page_detection_topic_agnostic} and show that models trained on SEO features consistently outperform those trained on TAG features. The SEO GBDT model outperforms TAG GBDT accuracy and recall by 14\%, achieving 96\% accuracy, 0.96 F1. Furthermore, the TAG paper reports an accuracy of 83\% for a linear SVM classifier \citep{page_detection_topic_agnostic}, which our linear SVM outperforms by 10\%.

The TAG dataset assigns domain-level labels to article-level, content-based features, so every article from a given domain has the same label, but different features. On the other hand, SEO features exist at the domain level---each article is assigned the SEO features of its domain, and so every article from a given domain has the same label and the same SEO features. As the sample of articles in the PoliticalNews dataset is drawn such that it is representative of the domain labels \citep{page_detection_topic_agnostic}, it is not surprising that domain-level SEO features outperform TAG features on these labels.

Of the flat models trained on the MBFC* dataset, GBDT consistently yielded the highest F1 on the reliability and absolute bias tasks and was comparable to Random Forest on the relative bias task. Performance on MBFC* is understandably lower than for PoliticalNews, as it's domain list is an order of magnitude larger and so the task is more challenging. This is also shown by SVM's success on PoliticalNews (SEO F1 = 0.93) vs. its failure across all three tasks on MBFC* (F1 $<= 0.01$).

\begin{table*}[!h]
  \centering
  \begin{tabular}{ccccccccccl}
    \toprule
    &  \multicolumn{4}{c}{\makecell{PoliticalNews}} & \multicolumn{6}{c}{\makecell{MBFC*}} \\
        \cmidrule(lr){2-11}
     &  \multicolumn{2}{c}{\makecell{SEO}} & \multicolumn{2}{c}{\makecell{TAG}} & \multicolumn{2}{c}{\makecell{Reliability}} & \multicolumn{2}{c}{\makecell{Abs. Bias}}  & \multicolumn{2}{c}{\makecell{Rel. Bias}} \\
    \cmidrule(lr){2-11}
    &Acc&F1&Acc&F1&Acc&F1&Acc&F1&Acc&F1\\
    \midrule
    GNN$_{to}$ & - & - & - & - & 84 & \textbf{.82} & 85 & \textbf{.79} & 70 & \textbf{.69} \\
    GNN$_{sb}$ & - & - & - & - & 82 & .79 & 83 & .79 & 66 & .62 \\
    \midrule
GBDT & 96.2 & \textbf{0.96} & 82.5 & \textbf{0.82} & 83.5 & \textbf{.76} & 83.7 & \textbf{.65} & 71.4 & .65 \\
    RF & 91.3 & 0.92 & 78.3 & 0.76 & 82.2 & .74 & 82.9 & .63 & 71.5 & \textbf{.66} \\
    DT & 93.6 & 0.94 & 80.7 & 0.8 & 77.0 & .69 & 76.5 & .56 & 63.5 & .59 \\
    MLP & 81.0 & 0.81 & 82.1 & 0.81 & 61.4 & .35 & 62.5 & .14 & 54.6 & .22 \\
    SVM & 92.7 & 0.93 & 80.1 & 0.8 & 63.3 & .01 & 73.2 & .00 & 55.7 & .00 \\
  \bottomrule
\end{tabular}
\caption{Comparison of flat and graph-based detection methods on each of the MBFC* and PoliticalNews \citep{page_detection_topic_agnostic} domain lists. For MBFC* we report F1 and accuracy on three sets of labels. For the PoliticalNews reliability labels, we compare the performance of models trained on the original TAG dataset (article-level) and our SEO dataset (domain-level).}
\label{tab:baseline_experiments}
\end{table*}

\begin{figure*}
  \includegraphics[height=4cm,width=\textwidth]{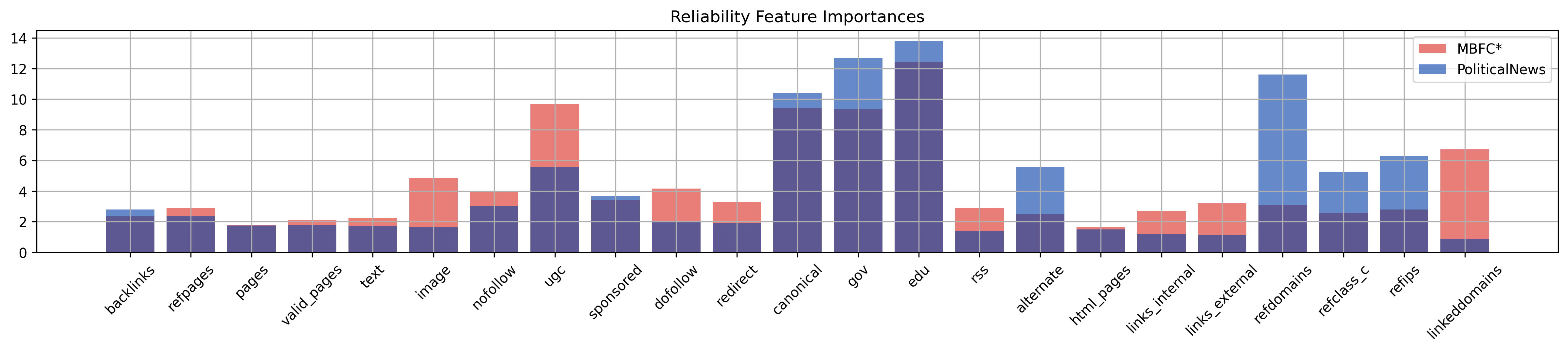}
  \caption{SEO attribute importances from Ahrefs on predicting reliability labels on the PoliticalNews and MBFC* datasets. For a full description of features used, see \href{https://ahrefs.com/api/documentation/metrics-extended}{Ahrefs.com}.}
  \label{fig:feature_importances}
\end{figure*}

The GNN models significantly outperform the flat classifiers on every task across a range of link networks and weight schemes. $GNN_{back}$ and $GNN_{out}$ are the GNNs trained with \textit{backlink} and \textit{outlink} weight normalization and schemes. Due to the smaller dataset size, the GNN methods are not suitable for the PoliticalNews dataset. However, similar GBDT feature importances are seen for both datasets on the reliability tasks, suggesting that classifiers trained on SEO features have generality (\autoref{fig:feature_importances}).


%

\subsubsection{Webgraph Experiments}

\begin{figure}[!htbp]
    \centering
    \includegraphics[height=6cm]{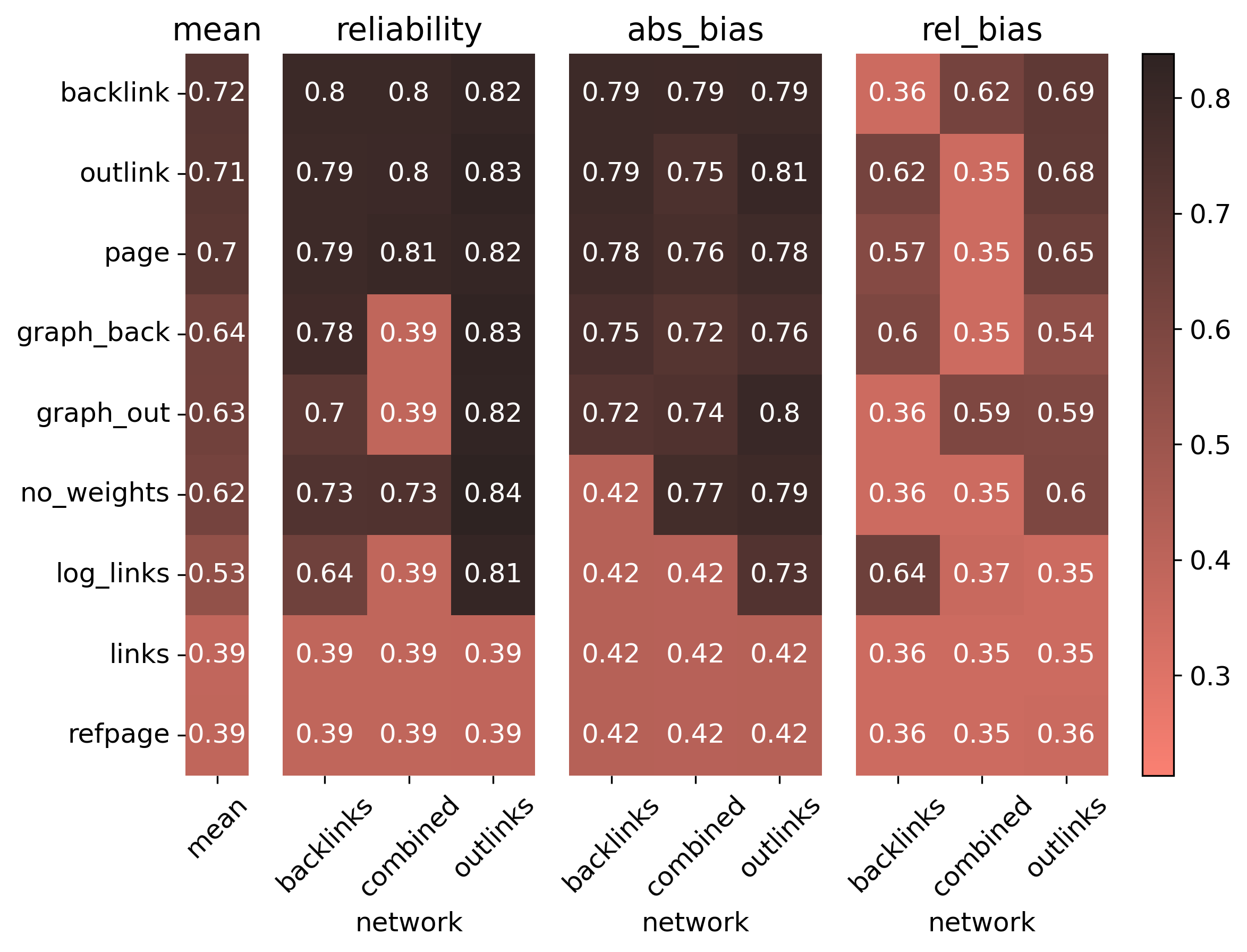}
    \caption{F1 scores for webgraph design space exploration, using GCN with various network structure and link weighting approaches.}
    \label{fig:network_experiments}
\end{figure}

Figure \ref{fig:network_experiments} provides F1 scores for each combination of network structure and link weighting approaches on the established detection tasks. We observe that the outlink network achieves the strongest overall performance. The combined network often results in the lowest performance, particularly on the relative bias task. Additionally, we observe that link normalization approaches have a substantial impact on model performance. The weighting approaches in Figure \ref{fig:network_experiments} are sorted in descending order from highest (top) to lowest (bottom) by the mean F1 score across all experiments. The backlink weighting achieves the highest mean F1 score. Interestingly, the highest F1 score on the reliability task is obtained by the outlink network with no link weights. However, in general the decision to normalize is more important than the exact normalization method employed, and further performance increases with more appropriate weighting approaches have smaller gains. It is also worth noting that the graph normalized variants, with sampled degree matrices $\hat{D}$, underperform compared to the backlink and outlink weightings.

\begin{figure}[!htbp]
    \centering
    \includegraphics[height=6cm]{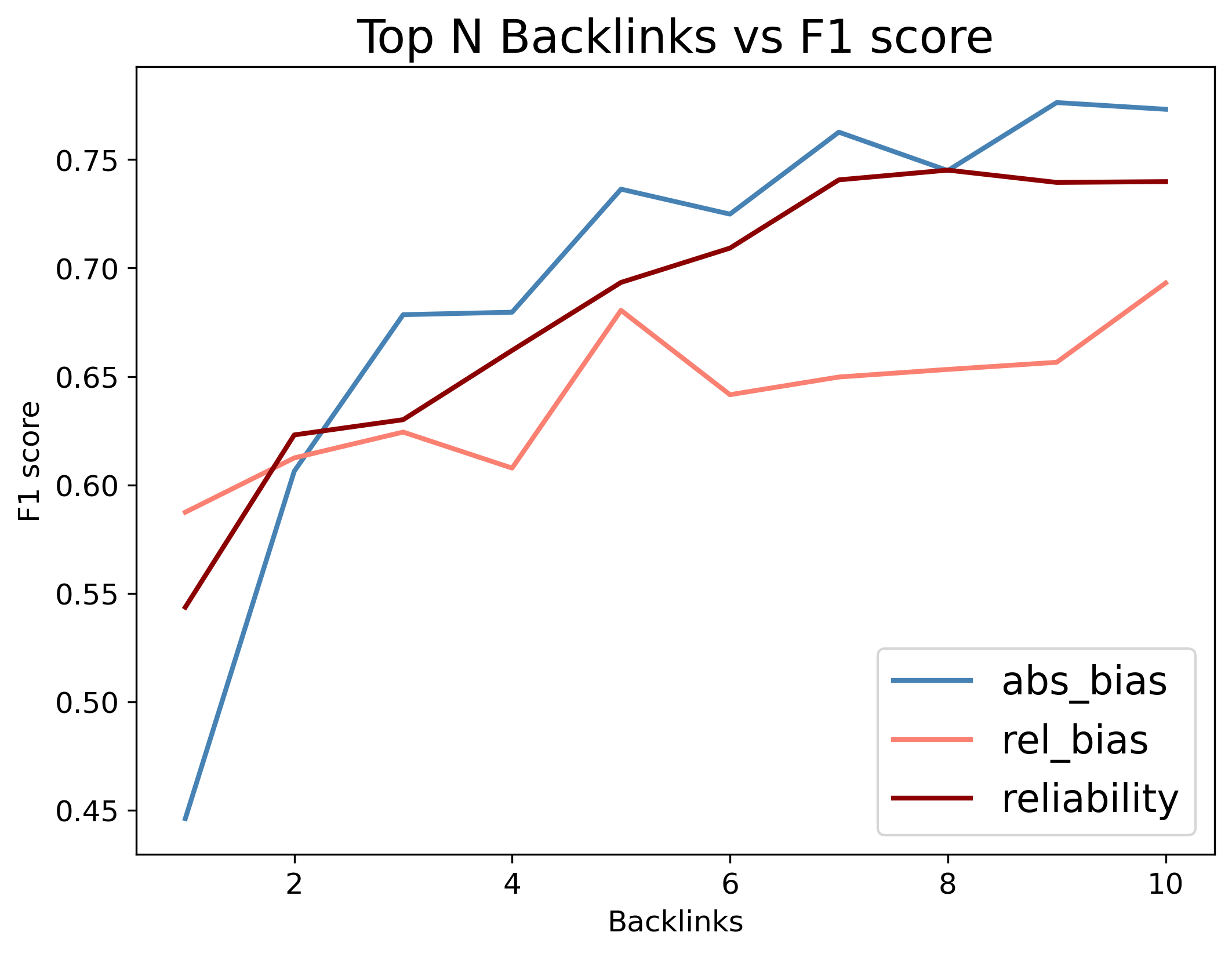}                       
    \caption{Increasing backlink context improves the performance of our models, to a point.}
    \label{fig:topn_backlinks_experiments}
\end{figure}

We find the inclusion of webgraph and SEO context is strongly related to the overall performance of our models. Figure \ref{fig:topn_backlinks_experiments} shows the effects of including a varying number of backlinks in the backlink network across each task (the top N backlinks, varying between 1 and 10). We use log weights, as we found those weightings to be the most intuitive as we vary N. We observe that, for all tasks, as the number of backlinks increases, performance tends to increase as well. Additionally, as we increase the number of backlinks used we observe diminishing returns. The F1 score plateaus at 0.76 with top 7 backlinks for the reliability task, 0.78 with 9 backlinks for the absolute bias task, and 0.68 with 10 backlinks for the relative bias task.


\subsubsection{Black-Hat SEO}
We find evidence that classic black-hat SEO techniques are being used disproportionately in support of unreliable news sites. Figure \ref{fig:blogping} in the appendix shows the negative correlation between reliability and percentage of backlinks coming from blogspot domains ($R = -0.25$). This is indicative of "Blog Ping", where link farming---an SEO tool to generate backlinks---is done on domains hosted by top blogging sites, because such sites are known to quickly attract web crawlers for indexing \citep{black_hat_seo_review}. Similarly, black-hat link building is often done through user generated content---namely forums and comment sections \citep{black_hat_seo_review}. We observe similar negative correlations between the `ugc' user-generated content attribute \& reliability ($R = -0.1$).

\section{Misinformation Source Discovery}

\begin{figure}[!htbp]
    \centering
    \includegraphics[width=6cm]{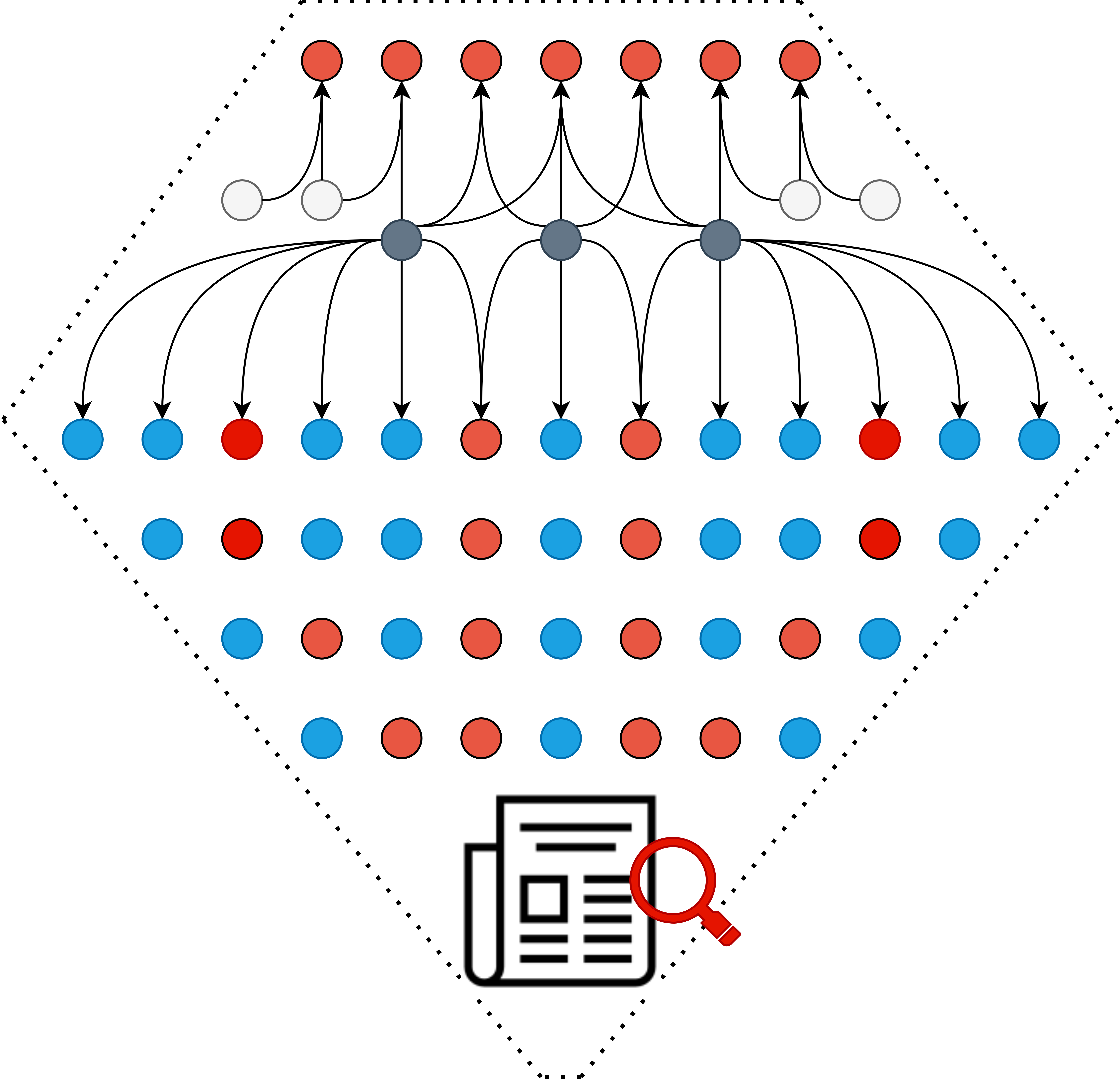}                       
    \caption{Diagram of the multi-step domain discovery process. We begin with a list of known unreliable domains (red)(1), then analyze the backlink network (white) to uncover link schemes (grey)(2), and pull the outlinks (blue) for these link schemes (3). To filter out noise, we apply a news classifier (4), a bias classifier (5), and a reliability classifier (6). Finally, we validate the results.}
    \label{fig:domain_discovery}
\end{figure}

Given the webgraph of a labeled seed list, we seek to capture a subset of nodes which maximizes the likelihood of nodes within that subset being unreliable. As described in \autoref{fig:domain_discovery}, our method is a two-part process based upon the hypothesis that backlink sites that heavily link to known misinformation sources also link to unknown sources. 



\subsubsection{Link Scheme Identification}
In the first stage of the discovery process, we attempt to capture backlinking domains with suspicious linking patterns. We extract the subset of backlinking sites that link to multiple sites labeled as unreliable (breadth constraint, $\beta$). We further parameterize by the total number of backlinks targeting unreliable sites (depth constraint, $\alpha$). Algorithm \ref{alg:link_scheme_identification} details this process.

\begin{algorithm}
\caption{The link scheme identification algorithm}\label{alg:link_scheme_identification}
\begin{algorithmic}
\Require $\beta_{min} \geq 0$ \Comment{Breath criteria}
\Require $\alpha_{min} \geq 0$ \Comment{Depth criteria}
\State Input: $G \gets (V,E)$       \Comment{Backlink Network}
\State $\mathcal C \gets []$    \Comment {Link Scheme Candidates}

\ForEach {$(s, t) \in \mathcal G.edges()$}
    \If{$t.reliability = 0$ } \Comment {Unreliable News Sources}
        \State $C.append(s)$
    \EndIf
\EndFor

\State $L \gets []$       \Comment {Link Schemes}
\ForEach {$c \in \mathcal C $}
    \State $S \gets G.successors(c)$
    \State $\beta \gets \Sigma^{S}_{s}{1}$
    \State $\alpha \gets \Sigma^{S}_{s}{s.backlinks}$
    \If{$\beta \geq \beta_{min} \land \alpha  \geq \alpha_{min}$ }
        \State $L.append(c)$
    \EndIf
\EndFor

\end{algorithmic}
\end{algorithm}
\renewcommand{\labelenumii}{\theenumii}
\renewcommand{\theenumii}{\theenumi.\arabic{enumii}.}

\subsubsection{News Classification} 
To identify candidate news domains, we attempt to classify news domains in link scheme outlinks. Link scheme outlinks are noisy and can contain link farms, businesses, and other irrelevant sources. We train a classifier on SEO features to determine which link scheme outlinks are news sources. We draw positive samples from our MBFC* dataset and we draw negative samples randomly from the Common Crawl dataset \footnote{\url{https://commoncrawl.org/}}. We found that a GBDT with 50:50 negative to positive sample ratio obtained the best results, obtaining 89\% accuracy and 0.87 binary F1 score. Increasing the number of negative samples from common crawl had little effect.


\subsubsection{Reliability Classification}
To identify which news site candidates are unreliable, we use our best-performing flat classifiers. Specifically, we apply both the reliability and absolute bias classifiers. Although we expect the GNNs to outperform the flat baselines on the reliability task, inference with a GNN requires building a link network for the URLs on which we run inference. As the reliability classifier is trained only on domains with $>$10k backlinks, we use the same criteria to filter candidate domains because we expect the classifier's performance to be inconsistent outside this range. Finally, we manually evaluate the reliability of the candidates, using translation tools where necessary.

\subsection{Discovery Results}
We run our discovery process for two datasets, the PoliticalNews evaluation set \citep{page_detection_topic_agnostic} and our MBFC* dataset. Parameters along with resulting domain counts are provided in Table \ref{tab:discovery_params}.

\begin{table*}[]
\begin{tabular}{llllllllllll}
    \toprule
      & Seed & Backlinks & $\alpha_{min}$ & $\beta_{min}$ & $L$ & Outlinks & $MC$ & \& News & \& Misinfo & \& Biased \\
    \midrule
    PNews & 79   & 100       & 2.5k [1-10k]     & 2 [1-10] & 132          & 100 [10-100]      & 4.7k       &  3.1k       &   1.7k                &   1.4k         \\
    MBFC* & 1179 & 10        & 400k     & 3 & 270          & 100      & 11.6k      & 6.9k    & 3.4k              & 2k      \\
    \bottomrule
\end{tabular}
\caption{Parameters used for link scheme identification in algorithm \ref{alg:link_scheme_identification} and number of candidates discovered using both the PoliticalNews evaluation dataset and our MBFC* dataset as seed lists in the discovery process. }
\label{tab:discovery_params}
\end{table*}

\subsubsection{Evaluation on the PoliticalNews dataset}
To evaluate the performance of our discovery system, we employ the partial F1 score metric as proposed by \citet{domain_discovery_social_media}. We find that the partial F1 of our webgraph-based system (0.28) is on par with the best reported partial F1 (0.29) of the content and social media based system outlined in \citet{domain_discovery_social_media}, with both systems achieving a partial F1 of ~0.25 for a large number of system configurations. The partial F1 metric defines the set of true positive unreliable domains as those labeled mixed reliability or worse by MBFC. We acknowledge that these results are not directly comparable as the current work uses a more up-to-date list of domain labels from MBFC. Rather, these results are meant to give an indication of performance for a system that is both challenging to evaluate and novel in terms of the methods used (i.e. webgraph vs content and social media based).

\begin{figure}[!h]
    \centering
    \includegraphics[height=8cm]{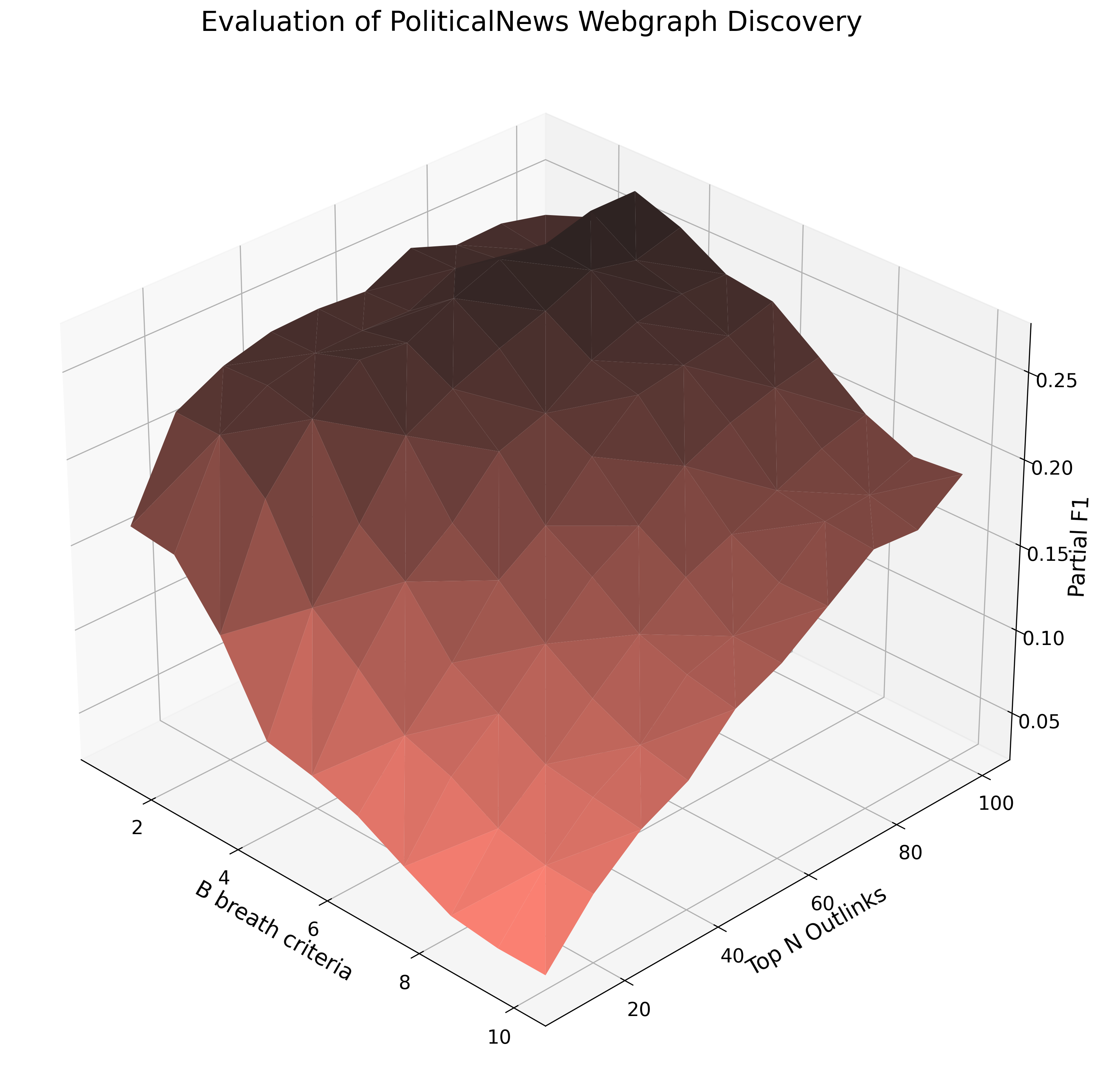}
    \caption{Link Scheme Identification; the x-axis represents the \# of outlinks used for each link scheme site, the z-axis shows the \# of misinfo sites linked to ($\beta_{min}$), and the y-axis shows partial F1 score \citep{domain_discovery_social_media}.}
    \label{fig:discovery_eval}
\end{figure}

Figure \ref{fig:discovery_eval} shows results from a design space exploration of the discovery algorithm. We find that optimal parameters with respect to partial F1 on the PoliticalNews dataset are $\alpha_{min} = 2500$, $\beta_{min} = 2$, and 100 outlinks per link scheme. Most notably, we find that our system is recall-bounded. That is to say that as we increase recall, by lowering the $\beta_{min}$ criteria (finding more link schemes) and increasing the number of outlinks used per link scheme, we increase the partial F1 score. In doing so, we observe diminishing returns, analogous to our GNN experiments showing how F1 increases with the number of backlinks used in the webgraph for reliability classification (figure \ref{fig:topn_backlinks_experiments}).


\subsubsection{System Demonstration on our MBFC* dataset}
In order to run the discovery process for the full list of domains in table \ref{tab:survival_rates}, we select parameters based on the optimal parameters from the design space exploration on the PoliticalNews dataset in figure \ref{fig:discovery_eval}. However, certain parameters are dependent on the size of the seed list; the depth constraint $\alpha_{min}$ represents the minimum number of links to unreliable domains required for a link scheme. Since our MBFC* dataset has an order of magnitude more seed domains, our criteria must increase to reflect this. In Table \ref{tab:discovery_params} we provide the parameters used in our discovery process. From left to right, we follow the order of execution in Algorithm \ref{alg:link_scheme_identification}: number of backlinks per known unreliable domain, $\alpha_{min}$ and $\beta_{min}$ link scheme criteria, $L$ link schemes, number of outlinks per link scheme, and $MC$ misinformation source candidates. Included are resulting domain counts after each stage in the discovery process from the size of the initial seed lists, to the final set of $MC$ misinformation source candidates that are classified as politically extreme, unreliable news domains. Brackets imply the range of parameters used in experimentation.

We observe evidence supporting the hypothesis of the discovery method: that link scheme sites which link to known unreliable sites also link to unknown unreliable sites. Table \ref{tab:misinfo_rates} provides the rate of unreliable domains in each explored list, as approximated by the reliability classifier. We find that link scheme outlinks contain a higher proportion of domains classified as unreliable than a random sample of 3241 domains from Common Crawl and the combined outlink and backlink networks. Further improvements are gained by filtering out sites that are not news according to our News Classifier.

\begin{table}
  \begin{tabular}{cl}
    \toprule
    Source & Misinfo Rate \\
    \midrule
    Common Crawl & 0.1\% \\
    Outlinks \& Backlinks & 10\% \\
    Link Scheme Outlinks & 29\% \\
    Link Scheme Outlinks \& News Sources & 49\% \\
  \bottomrule
\end{tabular}
\caption{Unreliable domain rates across the various domain lists as given by the reliability classifier}
\label{tab:misinfo_rates}
\end{table}

Ideally, the final stage of unreliable domain discovery would involve the manual validation of all candidates domains. However, due to time and budget constraints, we instead sample 10\% (200 domains) from the candidate list to estimate the accuracy of our discovery method. All labels were independently annotated by two researchers. Krippendorf’s alpha is computed to determine inter-annotator agreement. We find a strong agreement for reliability labels ($\alpha=0.86$), and some disagreement for bias labels ($\alpha=0.7$). As summarized in \autoref{tab:misinfo_evaluation}, we find that 47\% of domains frequently feature misinformation (mixed or worse), 37\% predominantly feature misinformation articles (low or very low), 46\% are frequently biased (not center), and 38\% predominantly feature politically polarized articles (extreme left or right). We note that 42\% of discovered domains are not news sources, which greatly impacts our results. However, given that a discovered website is a news source, it is highly likely to be unreliable (83\%), peddle known misinformation (65\%) or have extreme right political bias (65\%). The distribution of reliability and political bias labels for manually labeled news sources is given in Table \ref{tab:misinfo_evaluation}. 

Additionally, we profile the distribution of countries in our domain lists according to their IP addresses. For the MBFC* dataset, we find that 90\% of domain IPs are US-based. In comparison, we find that 80\% of the newly discovered domain IPs are US-based, which is confirmed by our evaluation (Table \ref{tab:misinfo_evaluation}). This reiterates our concern that misinformation source discovery systems are heavily biased based on their seed list of domains.

\begin{table}
  \begin{tabular}{cccccl}
    \toprule
     \multicolumn{2}{c}{\makecell{Reliability}} & \multicolumn{2}{c}{\makecell{Bias}} & \multicolumn{2}{c}{\makecell{Locale}} \\
    \cmidrule(lr){1-6}
    Label&\#&Label&\#&Country&\#\\
    \midrule
    V. Low & 63 & Ext. Left & 7 & US & 80\% \\
    Low & 10 & Left & 9 & EU & 15\% \\
    Mixed & 18 & Center & 14 & Middle East & 2\% \\
    High & 14 & Right & 6 & South Asia & 1\% \\
    V. High & 6 & Ext. Right & 67 & Misc. & 2\% \\
  \bottomrule
\end{tabular}
\caption{Evaluation of the discovered domains sample}
\label{tab:misinfo_evaluation}
\end{table}



We further find evidence of evasion and exit behavior in the operation of unreliable news domains. We observe domain switching through redirects and context switching through content changes. From our sample of candidate misinformation sources, 8\% are dead URLs, 4\% redirect to newer domains, and 2\% were once misinformation sources but have since `context switched' so that the domain is no longer a news site \footnote{We used archive.org to verify domains had context-switched.}. Together, 15\% of domains in our sample were previously news sites. In our annotations, we do not classify these domains as misinformation sources.


\section{Discussion}


\subsubsection{Survival Rates}
Given the ease of evasion and exit tactics like domain swapping, pivoting from news to satire, or shutting down altogether, more dynamic approaches for detecting and discovering misinformation sources are needed. Once unreliable news sites are exposed, they can be boycotted or penalized by search engines or social media companies. We find that 3 of 4 unmaintained domain lists published between 2019 and 2017 have domain survivability rates of 35-44\%.

\subsubsection{Webgraph GNNs}
The outlink network outperforms both the backlink and combined networks across the majority of our experiments (see Figure \ref{fig:network_experiments}). Intuitively we had expected backlinks to perform better, as they contain the link scheming behaviour that we use identify for site discovery. However, the better performance of outlinks can be understood in two ways; firstly, the information gained from the outlink network is denser due to there being fewer outlink sites than backlinks in general. If we were to compare results on a GNN trained on all backlinks vs. all outlinks (rather than 10 of each), we would likely get very different results. Unfortunately, such an experiment is not feasible given data limitations. Secondly, websites have more control over their outlinks than their backlinks, which may better provide a better proxy for site intent.

We find that performance on the absolute bias task, as measured by F1 score, is higher than for the relative bias task (Figure \ref{fig:network_experiments}). While this is in line with our hypothesis that link structure is more predictive of absolute bias, we note that it may also be due to the apparent skew in the unreliable labels towards the extreme right (Figure \ref{fig:bias_by_reliability}). This imbalance likely impacts performance on the relative bias task. 

Unsurprisingly, we find diminishing returns when adding more backlinking domains to the backlink network. 
As shown in Figure \ref{fig:topn_backlinks_experiments}, F1 score begins to increase more slowly as we add successive backlinks to the network. This is further supported by the design space exploration of the discovery process (figure \ref{fig:discovery_eval}), where increasing the number of outlinks used per link scheme improves recall, to a point.


\subsubsection{Discovery}
The core motivation for the development of the discovery process is that lists of unreliable domains will become quickly outdated. However the discovery process is still heavily biased towards the initial set of seed domains; Table \ref{tab:misinfo_evaluation} reveals that the known biases of locale and political orientation from the MBFC dataset (Figure \ref{fig:bias_by_reliability}) are reflected in the set of discovered domains; although slightly improved, the discovered domains are still predominantly US-based and skew extreme-right. We also note the potential political bias in using the partial F1 evaluation metric as it relies on the MBFC list.

We find that a small change in location distributions occurs during link scheme identification. This could allow researchers investigating unreliable news domains in a given region to set a starting locale distribution using both an initial seed list as well as by appropriately filtering the link scheme sites computed by Algorithm \ref{alg:link_scheme_identification}. While the danger of misuse is clear, we posit that our proposed methodology is extendable to other novel country and language contexts to better identify sources of misinformation. We further stress the language-, content-, and social media-agnostic aspects of the system in this regard.

We note that using archive.org in combination with our SEO classifiers is an effective way to investigate site history. As seen in our evaluation, 7\% of our sample is composed of non-news domains with scopes that have changed over time. However, backlink profiles are `sticky' and in many cases live longer than the actual content of the domain. By analysing mismatches between a domains link profile and its current content, we can uncover evasion and exit strategies.


\section{Limitations and Future Work}


We observe that blocklists of unreliable news domains have poor survival rates, with more than 50\% of such domains being parked or returning 404s to get requests within three years of publication. We employ a backlink heuristic to filter out 87\% of parked domains and 63\% of 404 domains that have less than 10k backlinks. This heuristic also filters out $>$500 domains that are still live, albeit with a minimal backlink profile. While the rationale here is to avoid training a model to predict dead domains as unreliable, it is still a limitation of the proposed method. As a result, domains should have a substantial history and a well-developed backlink profile for there to be confidence in the predictions of this detection method. Additionally, as with competing detection and discovery methods, a seed list of labeled domains is required.

Data collection through proprietary APIs can be costly. This cost constrains the validation of our discovery algorithm, which is why we use flat classifiers for the discovery process. However, we have shown that the GNN method is superior for reliability and bias prediction and so we expect that the accuracy of the discovery method would be higher using the GNNs for inference. We note that these methods are not `data-hungry' and achieve impressive performance with limited backlinks and outlinks (figure \ref{fig:topn_backlinks_experiments}, \ref{fig:discovery_eval}).

In our evaluation of the unreliable domain discovery process, we find that the lines between extreme political bias and reliability become blurred. While our proposed method is successful in creating a new sample of domains with a high prevalence of unreliable sites among them, the final evaluation is not a simple process. We encounter a broad range of domains in the link network, with news sites from all over the world appearing in the sample. In many countries and contexts, the spectrum of political bias may not exist on a clear left-right scale, so it is natural that inter-annotator agreement for bias labels is lower. We also encountered numerous satire sites which can be difficult to classify. Our aim is not to ignore the subjectivity of reliability and we acknowledge that the evaluation of candidate unreliable domains requires the development of thorough labeling protocols.

Finally, despite much of the related work taking advantage of social networks to determine news source reliability and to discover new unreliable domains, we exclusively look at webgraph and SEO attributes. In future work, we plan to combine our current webgraph approach with social network data and website content.



\section{Conclusion}
Progress in the misinformation space is typically realized by the publication of lists of domains that curate unreliable content and produce misinformation. However, due to the transient nature of many of these sites, we demonstrate that such lists do not stay relevant for long. With this in mind, we present a novel dataset crafted from a combination of webgraph data and SEO attributes. We provide strong GNN baselines for predicting news reliability and political bias labels on our MBFC* dataset, with F1 scores of 0.84 \& 0.81 respectively. We achieve SoTA results with an F1 score of 0.96 on the PoliticalNews dataset.

Additionally, we propose an algorithm for discovering new unreliable news domains. By leveraging signals of blackhat link-building methods and the natural homophily within webgraphs, we introduce a graph-based methodology to generate sets of candidate news sources. Applying the flat classifiers trained on our SEO dataset to these candidates, we effectively filter out false positives. Based on partial F1 scores, our content and social media agnostic algorithm is en-par with \citet{domain_discovery_social_media}. Evaluating this algorithm on the MBFC* dataset, 47\% of discovered domains are unreliable and 46\% are biased. We provide annotations for a sample of these newly discovered unreliable domains.

We demonstrate that webgraphs provide a strong signal for numerous tasks in the misinformation ecosystem. This methodology can allow online misinformation research in new country-contexts where social media usage may be low and English may not be the predominant language.

\section{Ethical Statement}
We acknowledge a growing distrust in misinformation detection systems as many feel that their advancement coincides with the erosion of free speech. While our unreliable domain discovery process is a crucial line of research to enable informed discussion about media reliability, it could also lead to misuse if applied to reliable or opposition domains by government censors.

A limitation of unreliable news lists is that they tend to contain more extreme-right domains than extreme-left domains (figure \ref{fig:bias_by_reliability}). While we do not know the true underlying left-right distributions of unreliable new sources, discovery algorithms trained on these imbalanced lists may bias the discovery algorithm disproportionately towards unreliable conservative domains. This is problematic as psychological research has found that both extreme-right and extreme-left individuals are susceptible to misinformation \citep{harper2019you}.

To validate the partisan concern, we show that a classifier can be trained to distinguish left-wing domains from right-wing domains. Consequently, discovery algorithms biased towards one side of the political spectrum may miss harmful domains that target the other. As a result, we propose the use of an `absolute' bias classifier for use in the discovery process, which is trained to identify politically extreme domains targeting both left and right.

\section{Acknowledgements}
We thank Isabel Murdock for developing the MBFC scraper. This work was supported in part by the Office of Naval Research grant (N000142112229) and the US Army grant (W911NF20D0002). Additional support was provided by the Center for Computational Analysis of Social and Organizational Systems (CASOS) at Carnegie Mellon University. The views and conclusions contained in this document are those of the authors and should not be interpreted as representing the official policies, either expressed or implied, of the Knight Foundation, Office of Naval Research, or the U.S. Government.

\bibliography{aaai22.bib}

\appendix
\begin{table}
  \begin{tabular}{cccl}
    \toprule
    Statistic & Backlinks & Outlinks & Combined \\
    \midrule
    Nodes & 14959 & 9827 & 20817\\
    Edges & 32110 & 30397 & 61917 \\
    Avg Degree & 4.29 & 6.18 & 5.95 \\
    Clustering Coefficient & 0.016 & 0.040 & 0.027 \\
    Characteristic Path Length & 0.401 & 0.140 & 0.334 \\
    Density & 1.4e-4 & 3.2e-4 & 1.4e-4 \\
    Network Assortativity & -0.104 & -0.033 & 0.100 \\
  \bottomrule
\end{tabular}
\caption{Summary Statistics for Webgraphs}
\label{tab:summary_stats}
\end{table}

\begin{figure}
    \centering
    \includegraphics[height=5cm]{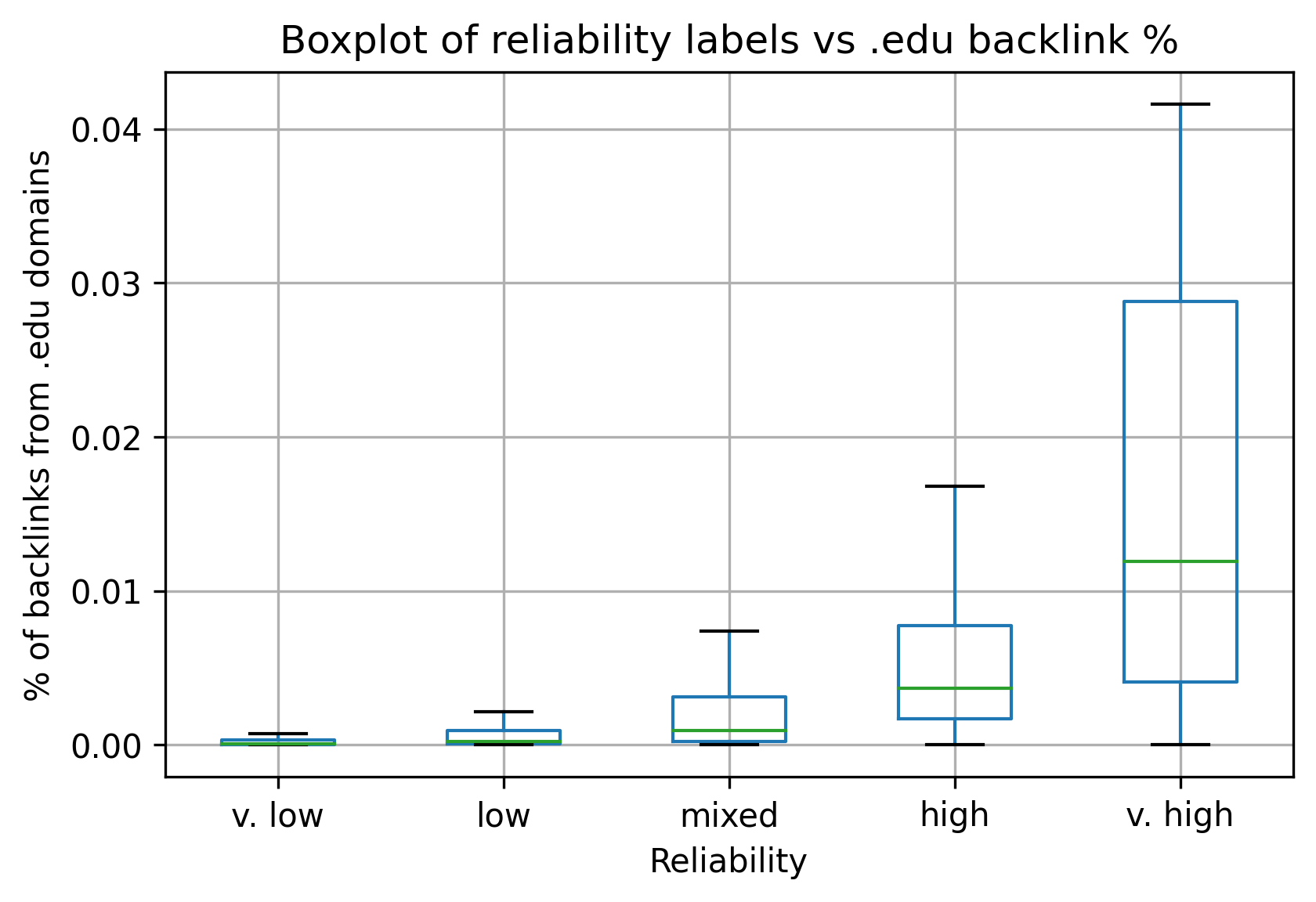}
    \caption{Percentage of backlinks coming from .edu domains, grouped by reliability.}
    \label{fig:edu}
\end{figure}

\begin{figure}
\centering
  \includegraphics[height=5cm]{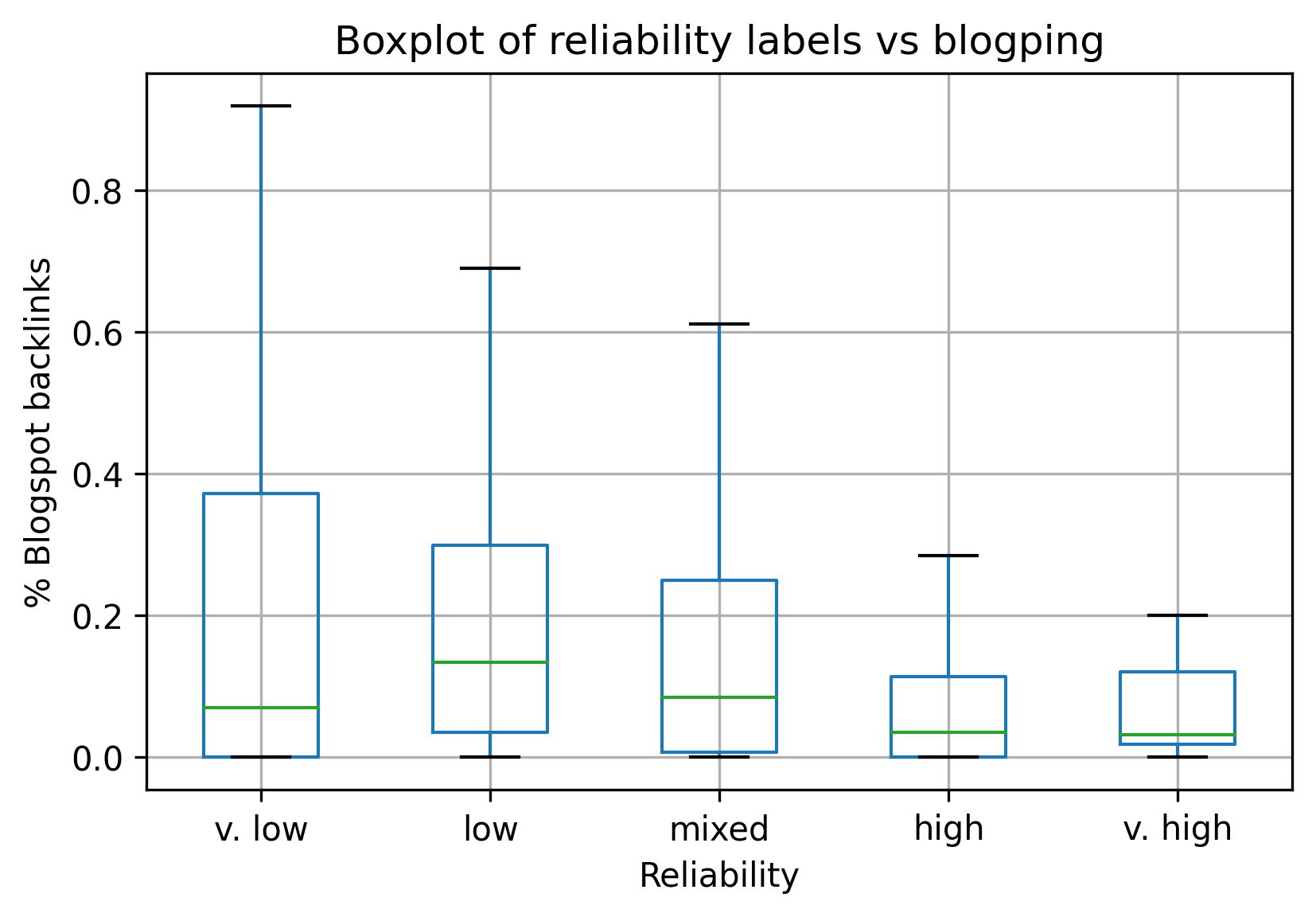}
    \caption{Percentage of backlinks from blogspot.com sites}
    \label{fig:blogping}
\end{figure}

\break
\section{Parked Domain Classifiers}

We evaluate two methods for classifying parked domains; a classifier trained on HTML \& HTTP response features \citep{parking_classifier} and a set of regular expressions (Regexps) that match known HTML patterns for parked domains \citep{regexp_parked}. The regular expression method does not require training. For the classifier, we construct positive labels by scraping 3k domains from sedo.com, a popular parking domain registrar. We construct negative labels by extracting random sample of 3k domains from Common Crawl.

Contrary to previous findings that a set of regular expression designed to catch parked domain HTML templates performs better than the trained classifier \citep{censor_lists}, we find that the regular expression method has low recall over our evaluation set of 3000 parked domains from sedo.com. Regexps obtains just 26\% recall on this list, with the trained classifier obtaining over 90\% recall. Performance between the two methods is compared in terms of accuracy and F1 score in Table \ref{tab:parked_domain}. 

We further find that Regexps have much lower recall over our news lists, with Regexps finding only 38\% of the parked domains that the trained classifier found. The status of these domains, once identified by the classifiers, was manually verified. We attribute the gap in recall to the regular expressions being outdated.

\begin{table}[!h]
  \begin{tabular}{ccccl}
    \toprule
    Method & Acc & F1 & \# in News List & Precision \\
    \midrule
    Regexp	                & 100\%	 & 0.43	 & 98 & 100\% \\
    Classifier	    & 95\%   & 0.92  & 253 & 96\% \\
  \bottomrule
\end{tabular}
\caption{Evaluation of Parked Domain Classifiers}
\label{tab:parked_domain}
\end{table}

\break

\end{document}